\documentclass{article}

\usepackage[left=2.5cm,top=2.5cm,bottom=2.5cm,right=2.5cm]{geometry}
\usepackage{graphicx,amsmath} \usepackage{lscape}
\usepackage{amssymb,amsfonts,amsmath}
\usepackage[usenames,dvipsnames]{color}%
\usepackage{graphicx}
\usepackage{cite}

\newcommand{\frad}{\emph{F.\ radicans}}%
\newcommand*\samethanks[1][\value{footnote}]{\footnotemark[#1]}

\begin{document}


\title{A neutral model can explain geographic patterns of sexual and
  asexual individuals}

\author{David Kleinhans \thanks{Corresponding author. Email:
    david.kleinhans@uni-oldenburg.de} \thanks{University of
    Gothenburg, Department of Biological and Environmental Sciences,
    G\"{o}teborg, Sweden} \thanks{University of Oldenburg, Department
    of Physics, Oldenburg, Germany.}  \and Daniel
  Johansson\thanks{University of Gothenburg, Department of Biological
    and Environmental Sciences, Tj\"arn\"o, Sweden} \and Lisa
  Sundqvist\samethanks[2] \and Ricardo T.\ Pereyra\samethanks[4] \and
  Per R.\ Jonsson\samethanks[4] \and Kerstin
  Johannesson\samethanks[4]}


\maketitle


\begin{abstract} Many species reproduce both sexually and asexually
  producing genetically unique and clonal recruits. The relative
  contribution of sexual and asexual reproduction is mostly considered
  a consequence of natural selection, for example, disfavouring sexual
  propagules during invasions. Here, we present a novel model for the
  invasion of species into a new habitat. The model is fully neutral
  with respect to the dispersal and the survival of sexual and asexual
  recruits. In addition to local dispersal through sexual and asexual
  reproduction, long-distance dispersal is implemented through a
  non-zero probability for long-range relocations. We parameterized
  our model using empirical data on the distribution of asexually and
  sexually recruited individuals of the recently established macroalga
  \emph{Fucus radicans} over its current area of distribution, the
  8000 year old Baltic Sea. As a solely stochastic alternative
  mechanism to natural selection, our simulation results suggest local
  abundance of one sex, that during a persistent phase of
  establishment will result in a heterogeneous distribution of sexual
  and asexual recruits. Specifically, the model suggests an initial
  general dominance of one or a few clones, and this pattern is
  observed in \emph{Fucus radicans} as well as in several other
  species with recent colonization histories. \end{abstract}


\section{Introduction}

The potential to reproduce both sexually and asexually, producing
genetically unique or clonal recruits, is widespread in protists
(including macroalgae) and aquatic and terrestrial plants, and also
found in some species of social and colonial animals such as aphids,
wasps, ants and corals \cite{Schon09,Vallejo10,Cousens}. Although
sexual reproduction implies extra individual costs it is likely to be
beneficial from an evolutionary perspective \cite{Otto09,Glesener78}
for the simple reason that more than 99\% of all species reproduce
sexually.

Fisher and Muller showed genetic disadvantages of clonal reproduction
in large populations being due to its weak evolutionary potential
\cite{Muller32,Fisher30,Park10}. This effect is particularly distinct
in non-equilibrium systems where adaptation to changing conditions is
favourable \cite{Otto09,Bengtsson00}. However, asexual reproduction
tends to dominate under certain circumstances, for example, during
invasions \cite{Kliber05,Kawecki08}, and at the edges of species'
geographic ranges \cite{Silvertown08,Kearney03} resulting in sexual
individuals and asexual clones showing spatial separation (termed
geographic parthenogenesis) \cite{Eckert03,Kearney05}.

Explanations of clonal advantage include better colonizing ability at
low densities \cite{Baker55}, adaptive mechanisms like the frozen
niche variation model \cite{Vrijenhoek84} and insusceptibility to
inbreeding \cite{Haag04}. In some species, asexual reproduction may
simply be the consequence of species formation by hybridization
\cite{Horandl09,Kearney05}. Peck et al.\ modelled range expansion of
competing sexual and asexual individuals into a habitat with a strong
environmental gradient and a severe selection pressure for local
adaptation \cite{Peck98}. The simulations confirmed the considerable
advantage of clonal reproduction in marginal habitats, under these
conditions. Other models show that asexual individuals may be favoured
in marginal populations because the relative cost of homozygosity will
be lower in already inbred, small and marginal populations
\cite{Haag04,Pujol09}. Although these models shed light on possible
mechanisms favouring an asexual reproduction strategy, they do not
address the more general question if facultative uniparental
reproduction per se may produce observed patterns of spatial
separation between sexual and clonal individuals. Hence, novel and
neutral models would be valuable for further investigations of spatial
population structures as a function of the interplay between sexual
and asexual reproduction.

The present contribution offers a novel approach to the problem of
understanding the role of asexual reproduction during invasions and
the consequences this may have on the population genetic structure of
species. We present a model and simulation results on invasions of a
species reproducing both sexually and asexually. A particular focus is
on the spatial distribution of sexes during invasion, which has not
explicitly been taken into account in this context before. As study
species we focus on the macroalga \emph{Fucus radicans}, a recently
formed species that has established in the 8000 y old Baltic sea
\cite{Pereyra09,Bjorck95} for which there is comprehensive population
genetic data \cite{Johannesson11}. We constructed an elementary model
that takes into account sexual and asexual reproduction, and
characteristics of long-range dispersal. Based on this model we
suggest a neutral interpretation of patterns of population genetic
structure that is a consequence of occasional long range dispersal and
an advantage of uniparental reproduction at low population
densities. More specifically, we show that the model can largely
explain spatial population structures of sexual and asexual
individuals observed in \emph{Fucus radicans}, and we argue that the
general framework is applicable to various species of terrestrial and
aquatic organisms. Since our model does not depend on natural
selection it may serve as a null-model in studies of geographic
parthenogenesis.

\section{\label{sec:material-methods}Material and
  Methods} \subsection{\label{sec:frad-baltic-sea} \emph{Fucus
    radicans} in the Baltic sea} The distribution of \frad, as
currently observed, is in the northern Baltic Sea (Gulf of Bothnia)
along the Swedish coast $60^\circ 20'$ to $63^\circ 25'$ N, and on the
Finnish coast $62^\circ 19'$ to $63^\circ 25'$ N, and in addition, the
species is found in two localities in Estonia (around $58^\circ 35'$
N) \cite{Johannesson11}. Further distribution south and east of the
present distribution around the Finnish coast, has been suggested, but
is not confirmed. The species lives in stands of densities from a few
to tens of individuals per m$^2$. The species was formed no more than
a few thousand years ago inside the 8000 y old postglacial Baltic Sea
and has undergone a recent (and perhaps, still ongoing) process of
establishment in the Baltic Sea \cite{Pereyra09}. As generation time
is relatively long in this species, the time since species formation
corresponds to no more than some hundreds of generations
\cite{Pereyra09}.

\emph{Fucus radicans} has separate sexes and gametes of both sexes are
released into the water during late spring to autumn, peaking at calm
days and full moon. Eggs are short-lived, have a negative buoyancy and
sink to the bottom close to the female individual. Indeed, preliminary
experiments in a closely related species in the field have confirmed
that sexual recruitment may only result in new attached individuals
within some meters distance of the mother individual
\cite{Serrao97}. Alternatively, new individuals may arise from budding
off small branches (adventitious branches) that are formed from the
main branches, and these branches reattach to the substratum through
the formation of rhizoids \cite{Tatarenkov05}. Adventitious branches
are only some $10\ \mathrm{mm}$ long and have a negative
buoyancy. Notably, also individuals recruited asexually produce
gametes and are sexually active (as males or females).

\begin{figure*}[tb] \centering
  \includegraphics[width=\textwidth]{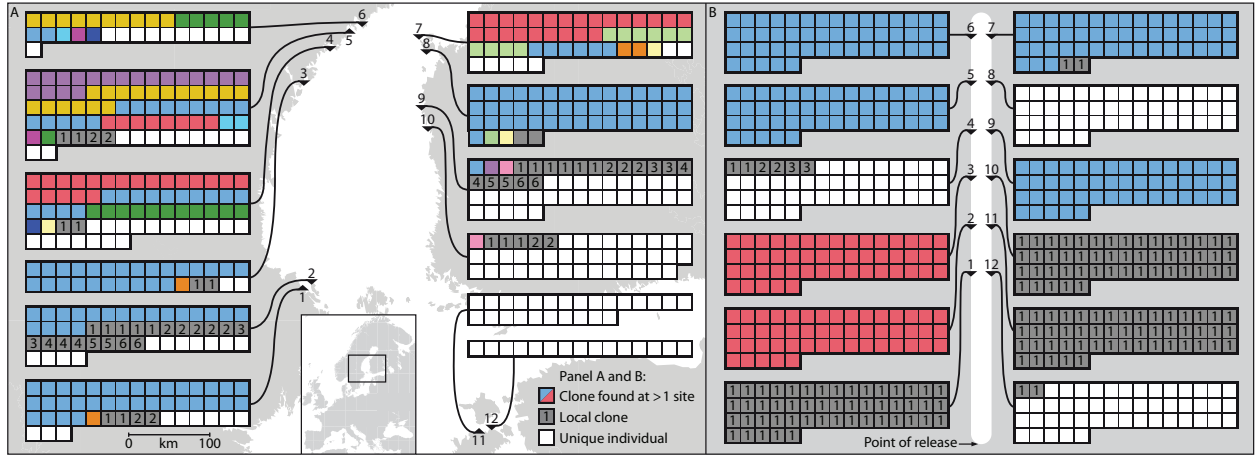}\\
  \caption{\label{fig:fucus-and-sim-distribution}Visualisation of the
    distribution of clones of \frad, both in data from empirical
    sampling campaigns (panel A) and in a representative realisation
    of our simulations (panel B, time of sampling and sampling
    locations are plotted in the center panel of Figure
    \ref{fig:SbS-threeV_twoRep}). In the panels the respective 12
    sampling locations are indicated on maps of the Baltic Sea (A) and
    an artificial circular habitat (B), respectively. The genetic
    composition of the sampled populations is illustrated graphically,
    where each square box in the respective populations refers to an
    individual. Boxes remain white if the individuals are unique, for
    instance since they directly originate from sexual reproduction or
    belong to a rare clone sampled only once in the entire sampling
    campaign. Clones encountered at several sampling locations (11
    pairwise different for the empirical sampling, panel A, and 2 for
    the simulated data, panel B) are indicated by a unique
    color. Clones unique to respective sampling locations are colored
    grey and numbered, with each number referring to a different
    clone. Please mind that local clones by definition are different
    between locations. Core features of the empirical data, such as
    the existence of local clones, of global clones at distant
    locations, and of unique individuals are reproduced with the
    simulations.}
\end{figure*}

Genotyping a large number of individuals using 5-9 microsatellite loci
has generated a rather comprehensive picture of the distribution of
individual multi-copy clones (hereafter “clones”), as well as unique
multilocus genotypes (hereafter “uniques”) the latter recruited by
sexual reproduction and zygote formation \cite[unpublished
data]{Tatarenkov05,Johannesson11}. The genetic structure is strongly
heterogeneous; the Estonian populations being completely sexually
recruited, some populations on the Swedish coast being almost
completely dominated by a widespread female clone, while other
populations in Finland and Sweden have a mixture of clones and uniques
(Figure \ref{fig:fucus-and-sim-distribution}). Although we here use
data on distribution of clones in \frad\ as our empirical example,
there are many other examples of species with both sexual and asexual
recruitment in which the pattern of distribution of clones looks very
similar to that of \frad, namely a mosaic pattern of asexually and
sexually dominant populations and an overall dominance of one or a few
clones (see Table 1 of the Supplementary Material for additional
examples, available on request).

\subsection{\label{sec:an-elementary-model}An elementary model
  involving sexual and asexual reproduction and long range dispersal}
The purpose of this study is to formulate and investigate an
elementary model of species invasion where individuals may reproduce
either sexually or asexually.  Both fertilisation of short-lived eggs
in the water and the negative buoyancy of adventitious branches make
\emph{Fucus radicans} a suitable species for efficient numerical
modelling, since we may assume that reproduction is a rather local
process. We intend explicitly not to involve adaptation to the
conditions of new environments during spread into marginal areas,
since the role of adaptation is not yet clear in \emph{Fucus radicans}
as in many other species. That is, genetic differences among
individuals are assumed to be neutral. In this respect our approach
offers an alternative to existing models for biological invasions in
marginal environments (see e.g. \cite{Peck98,Horandl09}).

From these basic considerations a model was constructed as follows: We
model the invasion of a species into a habitat consisting of $N$
patches of equal quality, where each patch can either be occupied by
one individual of known sex or be empty. The model is explicit in
space in a sense that the patches are arranged linearly on a circle
with periodic boundary conditions, where the distance $d(i,j)$ between
two patches $i$ and $j$ is defined as their shortest distance on the
circle (ranging from $0$ to $N/2$). Besides from being convenient for
technical reasons this configuration actually bears a resemblance to
the situation of \frad\ in the Baltic Sea (see below), which primarily
is located at a depth of $2$ to $8\ \mathrm{m}$ in coastal shallow
waters around the deeper basins of the northern Baltic Sea
\cite{Bergstrom05}. In this habitat we investigate a population
invasion model which involves two effects: 1.\ occasional long
distance dispersal of individuals through relocations\footnote{Here we
  use \emph{relocations} as a proxy for mechanisms for long distance
  dispersal, such as relocation of the whole organism, dispersal of
  fragments, buds or other asexually produced propagules.} of whole
organisms, which is known to have important effects during invasions
\cite{Johannesson88,Nichols94,Silvertown08}, and 2.\ a competition
between sexual and asexual reproduction, implemented in a way implying
as little bias as possible towards one of the modes of reproduction.

Getting a bit more specific we take advantage of the particular
features of reproduction in \emph{Fucus radicans}.\footnote{Generally
  the problem can be approached in a similar manner for other species
  by taking into account their respective routes of reproduction and
  dispersal.} Empty sites are colonised at rates depending on the
status of near-by sites. While the potential of colonisation through
asexual reproduction depends generally on the local density of
individuals only, the potential for colonisation through sexual
reproduction depends on the distribution of sexes in the neighbourhood
of site $i$. We assume that the local densities of female and male
individuals around patch $i$ are representative for the potential of
colonisation through sexual reproduction at this site. As an
intermediate step for each empty patch $i$ the local densities of
female, $p_f(i)$, and male, $p_m(i)$, individuals are calculated
as \begin{subequations} \label{eq:pf-pm-densities} \begin{eqnarray} \label{eq:pf-density}
    p_f(i)&=&\sum_{j=1}^N \left\{\begin{array}{ll}\mbox{individual at
          $j$ is female}&K(d(i,j))\\\mbox{individual at $j$ is
          male}&0\\\mbox{site $j$ is
          empty}&0\end{array}\right.\quad,\\ \label{eq:pm-density}
    p_m(i)&=&\sum_{j=1}^N \left\{\begin{array}{ll}\mbox{individual at
          $j$ is female}&0\\\mbox{individual at $j$ is
          male}&K(d(i,j))\\\mbox{site $j$ is
          empty}&0\end{array}\right.\quad. \end{eqnarray} \end{subequations}
Here, $K(d)$ is a kernel that models the potential of individuals to
contribute to the colonisation of a site at distance $d$. For reasons
of clarity and due to a lack of reliable empirical data on \emph{Fucus
  radicans} we at this stage assume that the dispersal capabilities of
gametes are sex-independent and equal to the dispersal capabilities of
asexual recruits through adventitious branches. For the kernel $K$ we
use a Gaussian-like localized function characterized by a scale factor
$\alpha$, \begin{equation} \label{eq:gaussian-kernel}
  K(d)=\mathcal{N}\exp\left(-\frac{d^2}{\alpha^2}\right)\quad
  \mbox{with}\quad\mathcal{N}=\sum_{d=-N/2}^{N/2}\exp\left(\frac{d^2}{\alpha^2}\right) \end{equation}
From the densities \eqref{eq:pf-pm-densities} per-patch colonisation
rates for sexual reproduction, $r_s(i)$, and asexual reproduction,
$r_v(i)$, at site $i$ are defined
as \begin{subequations} \label{eq:calculation-of-rates} \begin{eqnarray} \label{eq:rate-rs}
    r_s(i)&=&2\min(p_m(i),p_f(i))\quad \mbox{and}\\ \label{eq:rate-rv}
    r_v(i)&=&V\,\left[p_m(i)+p_f(i)\right]\quad. \end{eqnarray} \end{subequations}
Equation \eqref{eq:rate-rs} mirrors the fact, that for fertilisation
both female and male gametes need to be present at the same place at
sufficient probability. Here we assume that the limiting factor for
sexual reproduction is the minimum of $p_m$ and $p_f$, weighted by a
factor $2$ in order to equate the potential for sexual with the
potential for asexual reproduction at $V=1$ in the case
$p_f(i)=p_m(i)$ (see also \cite{Silvertown08}). The constant $V$ for
the asexual reproduction accounts for the differences in costs per
asexual propagule compared to the production of spawn, which can
result in different dispersal capabilities (for \emph{Fucus radicans}
we assume $V<1$, since the large amount of biomass of adventitious
branches suggests higher per recruit costs for asexual
reproduction). Technically $V$ can be interpreted as the rate of
successful asexual reproduction of one individual in an otherwise
empty habitat.

Individuals go extinct at a per individual extinction rate $e$. Long
distance dispersal is included through relocations of individuals at a
per-individual relocation rate, $r$. At the beginning of a relocation
the source patch becomes unoccupied. Then the destination patch is
drawn from a power-law kernel $K_r(d)=d^{-\beta}$ with $\beta\ge 0$
normalised to the size of the habitat accordant with Equation
\eqref{eq:gaussian-kernel}. In particular for small values of $\beta$
$K_r(d)$ models long-range dispersal \cite{Shaw95}. If the destination
patch is already occupied, the dispersing individual is
disregarded. Otherwise it occupies the destination patch.

\subsection{\label{sec:numer-invest-invas}Numerical investigation of
  the invasion process} We investigated the model numerically with the
intention to identify its key features. Instead of a complete
characterisation of the model, representative parameter values were
selected for discussion.  The model was implemented in continuous time
using Gillespie's algorithm for stochastic simulations
\cite{Gillespie77}. For characterisation of the genetic structure
during simulation each individual was labelled with a unique ID, which
was copied upon relocation and asexual reproduction. Individuals
originating from sexual reproduction were given a new ID. Our main
analysis builds on simulations performed on a habitat consisting of
$N=100\,000$ patches, which initially were all empty. At simulation
time $t=0$ a number of $100$ patches arranged side by side were
populated with unique individuals of alternating sex, $50$ females and
$50$ males. In order to be able to discuss the impact of this starting
configuration on the simulation results, additional simulations
initiated from $100$ individuals ($50$ of each sex) distributed
randomly on the entire habitat were performed, which are described in
detail in the Supplementary material (available on request) and which
henceforth is referred to as the \emph{random habitat} case.

Simulations for three different per-individual rates of asexual
reproduction ($V=0.002$, $0.020$ and $0.200$) and two different
extinction rates ($e=0.001$ and $0.000$) were performed. The different
parameter values for $V$ open up for a discussion of the ratio of
asexual to sexual reproduction, which is unknown in many relevant
systems. Testing $e=0$ and $e>0$ enables for an estimation of the
impact of extinction events on genetic patterns. A combination of a
starting configuration with a certain set of parameters $V$ and $e$
henceforth will be called a \emph{scenario}. For the local dispersal
and the relocation kernel $\alpha=10$ and $\beta=1.25$ were used,
which resulted in a well-balanced proportion of local reproduction and
relocation of individuals to long-range relocation events. The
relocation rate was fixed at $r=0.000001$ and simulations were
advanced for $100\,000$ time units, which corresponds to $1\,000$ mean
life times of individuals in the scenarios with $e=0.001$.

\begin{figure}[tb] \centering
  \includegraphics[width=.5\textwidth]{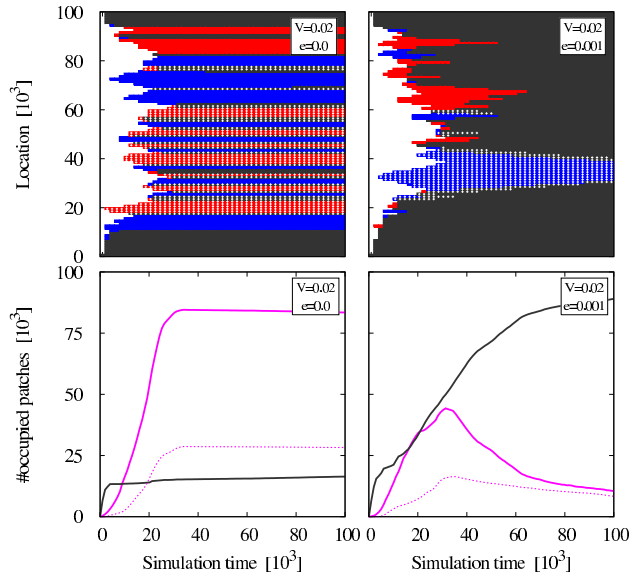}\\
  \caption{\label{fig:SbS-extinction} Realisations of two scenarios of
    the invasion processes for intermediate V=0.020 with extinction
    rates e=0.000 (left) and 0.001 (right).  The upper panels depict
    the sexual and genetic structure of the populations in the course
    of simulation time increasing from left to right. Please mind the
    circular form of the habitat implying that patch 100000 is a
    direct neighbour of patch 1. The colour coding illustrates the
    sexual and genetic constellation of groups of 1000 neighbouring
    locations in space and time: white areas indicate regions without
    any individuals and black areas populated areas with sexually
    reproducing individuals. In coloured areas only asexual
    reproduction is observed with all individuals being male (blue) or
    female (red). In addition to the sexual structure, the clonal
    structure of the entire population was investigated. Regions,
    where the dominating clone is found are covered by patterns of
    white crosses. The lower panels depict the respective amounts of
    individuals stemming from sexual reproduction (black) and asexual
    reproduction (purple, solid line) present in the entire habitat in
    the course of time. The dashed purple line displays the spread of
    the dominant clone.}
\end{figure}

\begin{figure*}[tb] \centering
  \includegraphics[width=\textwidth]{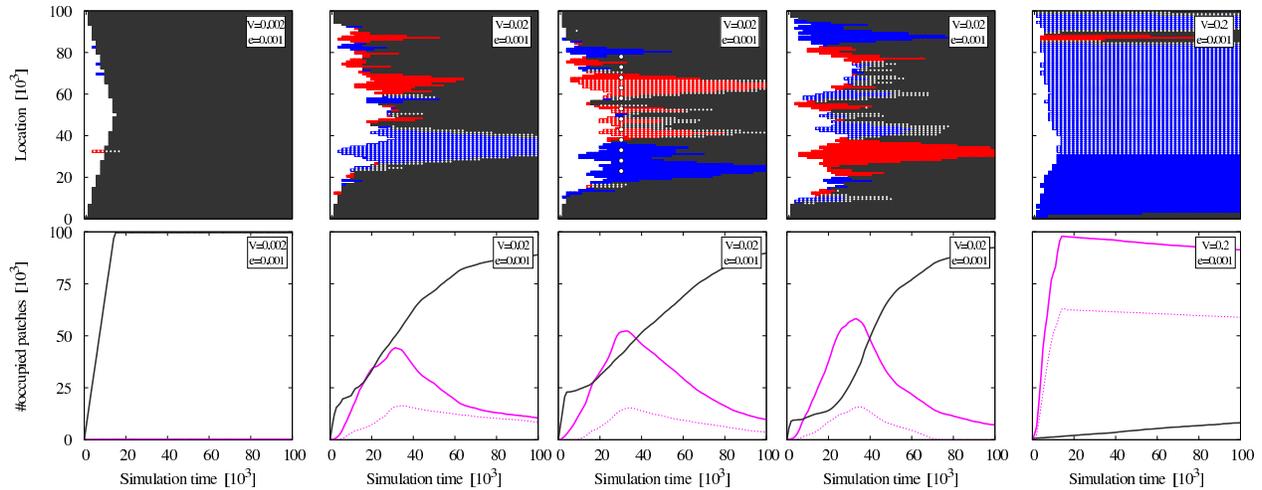}\\
  \caption{\label{fig:SbS-threeV_twoRep}Realisations of three
    scenarios of the invasion processes, each with extinction rate
    e=0.001.  The per-individual rates for asexual reproduction
    increase from V=0.002 in the left panel to V=0.200 in the right
    panel. The centre panels depict the first three out of ten
    replicates for the intermediate scenario with V=0.020. These
    realisations share the same parameters and starting
    conditions. Differences in the invasion process instead originate
    from the stochasticity intrinsic to the model. For details
    concerning the graphical presentation we refer to the caption of
    Figure \ref{fig:SbS-extinction}. The centre panel belongs to the
    replicate simulation used for preparation of the artificial
    sampling data depicted in Figure
    \ref{fig:fucus-and-sim-distribution}. The circles in this panel
    indicate time and location where the sampling was performed.}
\end{figure*}

\begin{figure}[tb] \centering
  \includegraphics[width=.75\textwidth]{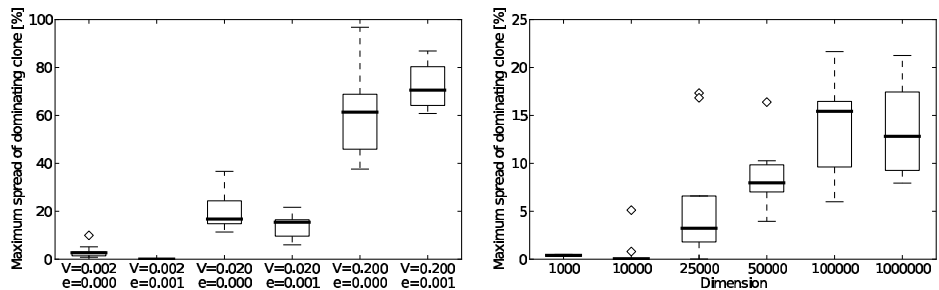}\\
  \caption{\label{fig:SbS-statistics}Statistical analysis of 10
    replicate simulations performed for each scenario (left panel) and
    for habitats of different size with V=0.020 and e=0.001 (right
    panel).  The box-and-whisker plots depict the maximum relative
    spread of the dominating clone in the course of the
    simulation. The whiskers indicate the lowest datum still within
    1.5 interquartile range, the diamonds outliers beyond this range.}
\end{figure}

For comparison with the empirical sampling results on the brown alga
\frad\ an artificial sampling procedure was performed on the results
of the numerical simulations: for this purpose at several simulation
times $12$ samples of $50$ individuals each were taken randomly
(without replacement) from distinct areas of the habitat (see white
circles in the centre panel of Figure \ref{fig:SbS-threeV_twoRep},
which indicate sampling time and locations) each consisting of $100$
neighbouring patches.

The state of the model is fully specified through the states of the
individual patches, the sexes of the individuals and their IDs, which
determine transition rates for the individual patches (e.g.\ for
extinction or relocation). Gillespie's approach implements the
individual transition rates in a stochastic simulation algorithm,
where events with higher rates in a self-consistent way occur at
higher probability in a certain period of time (and, hence, at a
higher frequency). As a consequence simulation results can differ
significantly between individual simulations. For each scenario
therefore $10$ independent stochastic simulations (henceforth called
\emph{realisations}) were performed in order to be able to quantify
their stochastic component. For determining the significances of
differences the results were subjected to an analyses of variance
(ANOVAs) with a significance level $0.05$. For the main analysis
$N=100\,000$ patches were used. In order to assure that our results
are not biased by the selected combination of parameters with this
particular habitat size additional simulations were performed for the
habitat dimensions $N=1\,000$, $10\,000$, $25\,000$, $50\,000$, and
$1\,000\,000$.

\section{\label{sec:results}Results} Figure \ref{fig:SbS-extinction}
depicts model realisations of two scenarios of the invasion processes
for intermediate asexual cost ($V=0.020$), that differ in extinction
rate. In the simulation shown in the left panel, extinction of
individuals was not considered explicitly (i.e.\ $e=0$). Starting from
$100$ occupied locations at the origin the invasion initially is
dominated by local sexual reproduction (see the initial increase to
$25\,000$ sexually reproducing individuals). Within short, however,
new colonies are founded at distant patches due to the long distance
relocation dispersal kernel.  New clonal colonies started from a
single individual consist of one sex only (left upper panel of Figure
\ref{fig:SbS-extinction}) since they were established by a single
asexually reproducing individual. In the absence of extinctions this
can result in a permanent separation of sexes. Although at $V=0.020$
the speed of local dispersal through sexual reproduction exceeds the
dispersal by means of asexual reproduction by more than one order of
magnitude, long-distance relocations followed by asexual dispersal
seem to be more effective in the long run in establishing populations
in new areas. After a short time colonies originating from relocations
and reproducing asexually prevent the sexually producing part of the
population from spreading through sexual reproduction by competition
for space. After the entire habitat is invaded at about simulation
time $40\,000$ the situation remains almost static and the sexes
remain separated until the end of the simulation. If a finite
extinction rate is considered as in the right panels, the genetic
patterns decay in the long run after peaking at about simulation time
$35\,000$. Instead, sexual reproduction slowly erodes the genetic
patterns generated by the invasion process. The increase in
individuals originating from sexual reproduction is almost linear
between simulation time $10\,000$ and $80\,000$. Note that the
presence of a finite extinction rate not necessarily seems to have a
strong influence on the maximum spread of the dominant clone in the
early phase of the invasion until simulation time $30\,000$ which is
indicated by the dashed purple lines in \ref{fig:SbS-extinction}).

Figure \ref{fig:SbS-threeV_twoRep} demonstrates the impact of the
per-individual cost for asexual reproduction, $V$, on the course of
invasions by comparing simulation results for three different
scenarios with different $V$. In the first case ($V=0.002$, left
panel) the potential for asexual reproduction is too low for playing a
relevant role in the invasion process. Interestingly the invasion
process is most efficient in this case with the entire habitat being
colonized in less than $20\,000$ time units. Large values for $V$ such
as $V=0.200$ (right panel) result in a big advantage of asexual
reproduction in range extension with a large potential for only a few
clones to dominate the entire population. In the long run sexual
reproduction, however, might be able to take over, as the lower right
panel suggests. An intermediate value of $V=0.020$ (centre panels)
allowed for a coexistence of sexual and asexual reproduction during
invasion. Simulations were performed by a stochastic simulation
algorithm implemented in continuous time, and this resulted in
realizations of the same scenario differing from one another. The
centre panels of Figure \ref{fig:SbS-threeV_twoRep} illustrate the
impact of stochasticity on the outcome of simulations by comparing the
course of invasions of three independent realizations performed for
the same scenario. These replicates differ significantly in the
general shape of the clonal fraction of the population in space and
time, in dominating sex and in the spread of the dominant clone.

Figure \ref{fig:SbS-statistics} contains a statistical analysis of the
maximal spread of the dominant clone in the six scenarios (left panel)
and of the impact of habitat size on the maximum relative spread of
the dominant clone (right panel). Generally the spreading potential of
the dominant clone increases with $V$, which confirm the main
observations from the Figure: for $V=0.020$ and $0.200$. Notably, the
presence of an extinction term $e=0.001$ has no significant impact on
the maximum spread of the dominant clone. Small habitat sizes reduce
the potential of a dominant clone to spread, which probably is due to
the advantage of dispersal through sexual reproduction in the early
phase of the invasion process (right panel of Figure
\ref{fig:SbS-statistics}). Between an habitat of size $100\,000$ and
an habitat consisting of $1\,000\,000$ locations, however, no
significant difference is observed with respect to the potential of
the dominant clone to spread in the population.

Empirical data on genetic structure for \frad\ populations show large
similarities with artificial samples from the model performed at
simulation time $30\,000$ such as dominance of a few clones, and a
strongly heterogeneous distribution of sexual and asexual individuals
(Figure \ref{fig:fucus-and-sim-distribution}).

\section{\label{sec:discussion}Discussion} Our elementary invasion
model can reproduce key features of spatial genetic structure in
\frad\ without invoking adaptations to geographically separated
environments. This is in contrast to earlier attempts to explain
observed patterns of spatial genetic structure, e.g. geographic
parthenogenesis, where most models have focused on explaining the
relative advantage of clonal reproduction in marginal
environments. Adaptive explanations of clonal advantage include better
average exploitation due to a diversity of ﾒfrozenﾓ genotypes
(Vrijenhoek 1984), insusceptibility to inbreeding (Haag and Ebert
2004, Pujul et al. 2009), avoidance of genetic recombination with the
central population (Peck et al. 1998) or as an indirect effect of
hybridity preceding asexual reproduction (H嗷andl 2009).

We here propose the present model as a null-model which any
adaptation-based hypothesis could be tested against. Our model
demonstrates that complex genetic patterns like the one observed,
e.g. in \frad\ could stem from rather basic effects during range
expansion in species for which both sexual and asexual reproduction is
possible. Standard genetic models, as well as models for invasion
processes, do not explicitly include aspects of the sexual structure
of populations. As a consequence it is difficult to consider both
sexual and asexual reproduction, since these routes of dispersal
generally depend on the sexual structure of local populations.

We addressed this requirement by developing a self-consistent basic
model that takes into account both sexes and their interactions for
modelling dispersal and invasion processes. Our modelling results
suggest that the genetic patterns observed in \frad\ might stem from
unlinked stochastic processes of the two sexes during the
establishment of this new species in the Baltic Sea. A similar
modelling approach is also applicable for a number of other species
with spatial genetic patterns resembling that of \frad\ (see Table 1
of the Supplementary Material, available on request). Consequently,
the dominance of a few clones needs not necessarily imply that these
clones are particularly well adapted but could instead originate from
the (partly stochastic) sexual structure of the population. For
dioecious organisms or outcrossing hermaphrodites sexual reproduction
is a rather local process. If an alternative route of dispersal
through asexual reproduction exists and occasionally individuals are
relocated to distant sites, asexual reproduction can outperform sexual
reproduction in the short run (see Figure
\ref{fig:SbS-threeV_twoRep}). In this case, genetic patterns during
invasion strongly depend on random effects of dispersal (compare the
replicates depicted in the centre panel of Figure
\ref{fig:SbS-threeV_twoRep}).

In particular, the dominant clone and its sex is a random phenomenon
varying strongly between realisations. If a finite extinction rate is
considered, in the long term sexual reproduction is able to take over,
if it is less costly for the individuals and, hence, more efficient
(Figure~\ref{fig:SbS-extinction}). If we view the competition between
sexual and asexual recruits for space as a competition for resources,
the short term handicap but long term benefit of sexual reproduction
is consistent with basic results from evolutionary game theory
\cite{Axelrod81} and with a recent modelling approach by Song et al.\
\cite{Song12}.

Results of a statistical analysis presented in Figure
\ref{fig:SbS-statistics} indicate, that small extinction rates and
also the limited size of the habitat used for the main analysis do not
have an impact on the simulation results. Concerning the spatial
patterns of populations after invasion we observe a local clustering
of sexes and clones (Figures \ref{fig:SbS-extinction} and
\ref{fig:SbS-threeV_twoRep}), which is consistent with previous
reports on the role of long-distance dispersal and local reproduction
\cite{Charpentier01,VanDerMerwe10,Kliber05}.

The structure observed significantly differs between realisations
(Figure \ref{fig:SbS-threeV_twoRep}, centre panels) and is a result of
stochastic effects during invasion. Just as in empirical observations
on \frad\ a single clone can spread through a great fraction of the
habitat and dominate the entire population (Figure
\ref{fig:fucus-and-sim-distribution}), which in case of the
simulations is a matter of chance rather than an individual advantage
of the dominant clone. Often the early phases of an invasion is of
particular importance for the long term genetic patterns, as these are
associated with \emph{priority} or \emph{founder effects}
\cite{deMeester02,Barrett08,Kliber05}. We therefore tested two
different starting configurations: In addition to the simulations
shown in Figures \ref{fig:fucus-and-sim-distribution} to
\ref{fig:SbS-statistics} a \emph{random habitat} case initiated from
$100$ individuals randomly distributed all over the habitat. Detailed
results prepared accordingly are available in the Supplementary
information (available on request). As a general trend, the
\emph{random habitat} starting conditions result in a much faster
invasion of the species, since the colonisation starts from a number
of different spots in the habitat. The potential for a single clone to
cover great parts of the habitat is strongly reduced. For this reason
we assume, that complex genetic structures with only a few dominating
clones as observed in \frad\ are likely to originate from an invasion
process (or speciation event) that started from a few spots instead of
from a great number of spots distributed over the habitat.

If we assume that the genetic structure in \frad\ indeed is driven by
the effects included in the simulation model, what patterns should be
expected in the future Baltic population of \frad? Based on our model
we expect that sexual reproduction will take over in the long run, and
the trend towards sexual dominance should start when the total
available habitat is colonized. When the rate of long-range
relocations is low the number of interfaces between areas dominated by
sexual reproduction is rather constant over time. Since we assumed
local Gaussian dispersal of sexual and asexual recruits, these
interfaces propagate at constant speed, resulting in a more or less
linear increase of the fraction of sexually reproducing individuals
with time. The linear relationship, however, is not likely to be found
in nature, since dispersal and relocation processes are more complex
and the restriction to one spatial dimension may be an unrealistic
simplification of many natural habitats. The qualitative effects of a
long-term success of sexual reproduction, however, can be expected to
hold. One single observation supports the basic mechanism of a
population being able to change sex-ratio rather dramatically over a
short period of time; a population of \frad\ sampled in 1996 outside
Ume{\aa} (Drivan) was found to be strongly female-biased at that time
($80\%$ females, N=82 \cite{Serrao99}), but re-sampling the exact same
locality in 2010 showed an almost equal sex ratio ($56\%$ females,
N=193, Pereyra et al., unpubl.).

The main effects of our model are likely to apply also to the broad
class of harsh marginal habitats subjected to periodic extinctions and
re-colonisations, e.g. during interglacial recolonizations
\cite{Kearney05,Johannesson11}. Based on these simulations we believe
that it is important to explicitly include the distribution of sexes
in future attempts to model the spread of species reproducing both
sexually and asexually. Moreover the fundamental mechanisms
implemented in this modelling approach should be investigated in
detail in future studies both in the laboratory and in field
experiments.

\section*{Acknowledgments} We are greateful for help with sampling of
Fucus radicans in Drivan by Ellen Schagerstr\"om, Lena Kautsky and
Angelica Ardehed, and in Finland by Jens Perus. This work was
supported by a Linnaeus-grant from the Swedish Research Councils, VR
and Formas (http://www.cemeb.science.gu.se).

\end{document}